\documentclass{article}

\usepackage[final]{neurips_2026_vericode}

\usepackage[utf8]{inputenc}
\usepackage[T1]{fontenc}
\usepackage{hyperref}
\usepackage{url}
\usepackage{tabularx}
\usepackage{booktabs}
\usepackage{enumitem}
\usepackage{amsmath,amsfonts,amssymb}
\usepackage{nicefrac}
\usepackage{microtype}
\usepackage{xcolor}
\usepackage{longtable}
\usepackage{pdflscape}

\title{When Code Works but the License Doesn't: Evaluating License Constraint Handling in Coding Agents\thanks{Accepted at the NeurIPS 2026 VeriCode Workshop.}}

\author{
  Sundaraparipurnan Narayanan \\
  AI Tech Ethics \\
  Bangalore, India \\
  \texttt{sundar.narayanan@aitechethics.com}
  \And
  Sandeep Vishwakarma \\
  AI Tech Ethics \\
  Bangalore, India \\
  \texttt{sandeep.vishwakarma@aitechethics.com}
}

\setlist[itemize]{leftmargin=*,nosep}

\begin{document}

\maketitle

\begin{abstract}
Coding agents increasingly perform repository-level software development tasks, including selecting dependencies, reusing and modifying source code, debugging implementations, and preparing software for deployment and distribution. These capabilities place autonomous agents directly within software supply-chain and licensing decisions that have traditionally been managed by developers, legal teams, procurement, and compliance processes. Existing evaluations of coding agents, however, primarily measure functional task completion and do not establish whether the resulting software is appropriate for its intended commercial use. We investigate whether coding agents can recognize, interpret, and preserve software-license constraints when constructing proprietary software intended for commercial use. We conduct controlled experiments across multiple coding agents and a diverse set of licensing and development conditions, including permissive and copyleft licenses, missing or contradictory licensing information, transitive dependencies, source reuse and modification, vendoring, dual licensing, source-available restrictions, deployment boundaries, and customer-distributed artifacts. We additionally introduce no-reference variants to examine whether agents autonomously discover licensing constraints when potentially problematic components are not explicitly identified. Agent traces, repository changes, dependency and source provenance, and final artifacts are evaluated against licensing ground truth using measures of functional success, license detection, commercial-use decision accuracy, incorporation-mode accuracy, provenance awareness, constraint preservation, final compliance, and escalation.

Our results indicate a persistent distinction between functional correctness and licensing correctness. Agents can successfully construct and test software while failing to establish licensing correctness, particularly when constraints involve transitive provenance, source modification, authorization, or changes in deployment and distribution context. Stronger behavior occurs when agents translate licensing information into concrete engineering actions, including dependency substitution, architectural changes, preservation of licensing information, or release gating. These findings suggest that software licensing in agentic development should be evaluated not as a license-classification problem, but as a contextual constraint-recognition and preservation problem across the software lifecycle.
\end{abstract}

\section{Introduction --- Context}

\paragraph{Agentic Coding Tools:} Coding tools with agentic capabilities have moved beyond single-shot code generation toward repository-level activity involving repository navigation, tool use, and code editing, execution and testing \cite{jimenez2024swebench,yang2024sweagent,zhang2024autocoderover,xia2024agentless,wang2024openhands}. These tools are capable of independently performing software-development tasks including identifying suitable libraries; accessing public repositories; selecting and installing dependencies; reusing and modifying existing code; resolving implementation problems; changing dependencies during debugging; preparing software for deployment and distribution. Understandably, their performance is not only influenced by the underlying model, but also the agentic pipeline, the harness, the agent-computer interface, available tools and the execution environment \cite{yang2024sweagent,xia2024agentless,wang2024openhands}. These tools enable a direct interaction between autonomous coding agents and the open-source software ecosystem in enterprise environment, while enterprise software development may have additional constraints including proprietary nature of the code; commercial distribution requirement with or without deployment as a commercial SaaS / service model. Open-source licenses or openly accessible code with commercial licenses (e.g BSL style) has varied conditions they impose on commercial use, modification, combination and distribution. This triggers a question as to whether an agent can produce a technically functional software while making an inconsistent or inappropriate licensing choice/ decision?

\paragraph{Motivation:} Our motivation for this study comes from the following 3 key premises:
\begin{itemize}
    \item \emph{License compliance is contextual:} Autonomous agents can introduce license risks through repository forking, dependency choices, source reuse, modification and packaging rather than only through generated text \cite{yang2024sweagent,wang2024openhands}. Hence, the consequence depends on the license, the technical relationship, modification, incorporation and distribution/deployment context by the agent \cite{pfeiffer2022licenseincompatibilities,xu2023pypiincompatibilities}. Further, dependency licensing is a graph-level problem because indirect dependencies can introduce obligations not visible from the top-level package \cite{pfeiffer2022licenseincompatibilities,xu2023pypiincompatibilities}. 
    
    \item \emph{The constraint-preservation problem:} In multi-agent workflows, a licensing constraint introduced at the beginning of a task may be lost during later stages such as delegation, implementation, testing, or release. This makes licensing a constraint-preservation problem throughout the workflow \cite{cemri2025mast}.

    \item \emph{Gaps in existing benchmarks:} Prior research indicates that code-generation capability does not necessarily imply accurate license awareness \cite{xu2024licoeval}. However, existing evaluations and benchmarks predominantly measure whether an agent completes a task or produces a patch that satisfies designated functional tests \cite{jimenez2024swebench,yang2024sweagent}. Recent benchmark work additionally motivates held-out, commercial, and long-horizon tasks to reduce contamination and better reflect enterprise settings \cite{deng2025swebenchpro}. Conversely, these measures do not necessarily establish whether the resulting software is appropriate for its intended commercial use. This leaves a gap between \textbf{functional task success} and correctness with respect to software-development licensing constraints. A recent survey covering more than 150 studies of LLM-empowered software-engineering agents organizes benchmarks around tasks such as code generation, translation, and repair, without identifying licensing or legal-compliance reasoning as a benchmark category \cite{guo2025llmagentsesurvey}.  
    \end{itemize}

In this study we examine whether coding agents can recognize and preserve license constraints when constructing commercially intended software. 

\paragraph{Why this premise is critical?} Software-license compliance is traditionally treated as a developer, legal, procurement or software-supply-chain responsibility, all of which traditionally are human-managed processes. Autonomous coding agents increasingly make decisions in this context that previously belonged to these human-managed processes. While agents may encounter licensing information at different stages of software development, the degree to which they can recognize licenses that are explicit or incomplete or contradictory or attached to a dependency or transitive dependence, or a copied source code or embedded in a vendored code or represented in README instead of appropriate license expressions needs evaluation. Further, commercial-use consequences depend not only on the license that is involved, but also how the software is used or whether code is modified or whether code is incorporated or how the end product is distributed or deployed over a network. Hence, license handling is not simply a license-classification task but a license recognition and preservation task and we wanted to evaluate it through this study. 

\section{Prior Research and Research Gap}
\paragraph{LiCoEval:} LiCoEval evaluates source similarity and license-information accuracy for LLM-generated code \cite{xu2024licoeval}. The research exhibits that the  code-generation capability and license-compliance capability can diverge, particularly for copyleft-licensed material \cite{xu2024licoeval}. However, the benchmark is function-level and does not evaluate autonomous dependency selection, repository modification, commercial deployment, obligation preservation or human escalation. Also the benchmark does not examine commercial-use decisions, explicit treatment of dependency/provenance chains; and persistence of license constraints across agents.

\paragraph{RepoComplianceBench: } RepoComplianceBench demonstrates that coding-agent compliance can be evaluated at the repository-trajectory level, including refusal, disclosure, verification and human handoff \cite{yang2026repocompliance}. This benchmark methodologically established that compliance can be evaluated as agent behaviour rather than inferred from the final code alone. However, the focus of this benchmark is governing rules concern AI contribution policies rather than the interaction of license type, modification, incorporation, deployment and distribution.

\paragraph{SOPBench: } SOPBench exhibits how natural-language procedures and constraints can be translated into structured action and verification functions, providing a useful template for executable licensing oracles \cite{li2025sopbench}. However, it does not evaluate OSS licensing, dependency-license compatibility, provenance or commercial-distribution obligations.

\paragraph{SWE Bench and other coding benchmarks:} Existing coding-agent benchmarks focus predominantly on functional outcomes \cite{jimenez2024swebench,yang2024sweagent}. However, functional test completion is not a reliable representation for license compliance. Also existing licensing evaluation is largely function-level and does not capture repository-scale autonomous decisions \cite{xu2024licoeval}. While, dependency research establishes direct and transitive incompatibilities, it does not, however, evaluate whether coding agents identify and remediate them during execution \cite{pfeiffer2022licenseincompatibilities,xu2023pypiincompatibilities}. Repository-level benchmarks can also suffer from test-oracle defects, specification ambiguity, and solution leakage \cite{jimenez2024swebench,deng2025swebenchpro,wang2025solvedissuesswebench}; however, the existing work does not fully examine package verification, untrusted repository instructions, copy-based reuse, or licensing obligations across complete agent workflows \cite{spracklen2025packagehallucinations,eliav2026knowyouragent,jahanshahi2025beyonddependencies}.
\paragraph{Research Gap: } Existing work therefore does not resolve whether an autonomous coding agent can recognize and maintain licensing constraints from task intake through dependency selection, source reuse, modification and commercial release. This study covers lifecycle-aware repository evaluation with contextual commercial-license reasoning, provenance, dependency closure, constraint preservation and appropriate deferral.

\paragraph{License Scope Relevant to the Study:} This study anchors on commercial licensing and release with a focus on licenses relevant to \textbf{commercial enterprise software construction} as referred to in \hyperref[tab:license-context]{Table~\ref*{tab:license-context}} and not every software license. Also, we take into account license evidence at repository and file level, as prior research suggest that top-level declaration can omit or conflict with licenses embedded in source files \cite{wolter2023licenseinconsistencies}.

\begin{table}[h]
\centering
\small
\begin{tabularx}{\textwidth}{@{}p{0.16\textwidth}p{0.20\textwidth}X@{}}
\toprule
\textbf{License} & \textbf{Broad category} & \textbf{Commercial-use relevance} \\
\midrule
MIT & Permissive & Generally permits commercial use; preservation of copyright/license notice remains relevant \\
BSD-2/BSD-3 & Permissive & Generally permits commercial use; notice and other license conditions remain relevant \\
Apache-2.0 & Permissive & Commercial use generally permitted; notice, attribution, modification and other conditions relevant \\
MPL-2.0 & Weak copyleft & Commercial proprietary software can incorporate MPL components, but obligations can attach to covered files/modifications \\
LGPL & Weak copyleft & Commercial proprietary applications can use LGPL components, but obligations depend on use, modification, linking and distribution \\
GPL & Strong copyleft & Commercial distribution can create significant licensing/source-availability consequences depending on how software is combined and distributed \\
AGPL & Network copyleft & Particularly relevant where modified software is provided through network interaction \\
SSPL & Source-available/restrictive & Offering the software as a service can trigger additional source-availability obligations under its terms \\
No license & No explicit permission & Public availability does not by itself establish permission for commercial reuse \\
Dual/commercial license & Multiple licensing paths & Agent may need to identify and select the appropriate licensing option for commercial use \\
Business Source License (BSL) & Source-available/restrictive & May impose production-use or time-based restrictions; source availability does not itself establish permission for the stated commercial use. \\
Elastic License 2.0 & Source-available/restrictive & May impose restrictions relevant to managed-service, commercial or customer-facing use; the applicable deployment and distribution context must be considered. \\
\bottomrule
\end{tabularx}
\caption{Licensing dimensions and commercial-use considerations evaluated in the benchmark.}
\label{tab:license-context}
\end{table}

\paragraph{Research Questions: } We structure our study with the following 4 research questions: 
\begin{itemize}
    \item \emph{RQ1. Licensing reasoning across the software lifecycle: } Can coding agents correctly identify, interpret, and preserve open-source licensing constraints when autonomously building software intended for proprietary/commercial use?
    \item \emph{RQ2. Contextual licensing decisions: } Does an agent's licensing decision remain correct when the way the code is used, modified, incorporated, deployed, or distributed changes?
    \item \emph{RQ3. Recognition versus contextual reasoning: } Do agents connect licence, modification, incorporation, deployment, distribution and  obligations, rather than recognizing the licence name alone?
    \item \emph{RQ4. Provenance and constraint preservation: } Do agents preserve licensing constraints when code is copied, modified, vendored or incorporated, and when the deployment or artifact boundary changes?
\end{itemize}

\section{Methodology}
We designed a controlled benchmark to evaluate whether coding agents can recognize, interpret, and preserve software licensing constraints during repository-level software development intended for proprietary and commercial use. 

\paragraph{Benchmark premise and decision context:} Commercial use itself is generally not prohibited by mainstream OSS licences and the relevant obligations depend on the licence together with modification, integration, distribution, deployment, artifact type, and notices. Hence, the benchmark is constructed around the observation that licensing consequences are contextual rather than determined solely by a license label. The benchmark includes permissive licenses such as MIT, BSD, and Apache-2.0; weak and strong copyleft licenses including MPL, LGPL, and GPL; AGPL; cases with no explicit license or contradictory licensing evidence; dual-licensing conditions; and source-available or restrictive licenses including BSL, Elastic License 2.0, and CCL. These categories are not treated as binary classes of 'commercially allowed' or 'commercially prohibited', but to have contextual considerations that are dependent on modification, incorporation mode, deployment context and distribution model among others. The license families considered in the experiments are summarized in \hyperref[tab:license-context]{Table~\ref*{tab:license-context}}.

\paragraph{Experimental design:} We conducted controlled experiments comprising 33 test cases across Codex, Cursor, and Antigravity. Case identifiers use T- and NT-prefixed labels for referenced/contextual cases and N-prefixed labels for no-reference variants; A/B suffixes identify paired conditions. The appendix taxonomy summarizes the scenario families. The no-reference cases were designed to evaluate whether agents independently surfaced licensing concerns without being explicitly directed to audit a particular component. Their outcomes were interpreted in relation to the observable development trajectory. Where possible, we distinguished failure to discover the relevant component or licensing evidence from failure to interpret an identified licence and failure to preserve an identified obligation in the final artifact. These stages were not treated as interchangeable, since a failure at discovery does not by itself establish that the agent would have failed to reason correctly if the relevant evidence had been available.

Each task specifies a functional software requirement and a commercial/proprietary intended use as context. This includes paired or near-replicate cases where one licensing-relevant condition changes while other task requirements remain substantially constant. We used fixed repository versions/commits in instances where we provided repository as a reference in the test cases. There were varying degree of relationships between the provided third party repo as reference and the software development task, covering licensing evidence, dependency relationship, source provenance, modification, incorporation mode, deployment context, distribution model, artifact boundary, and human intervention or authorization requirements. We used the default model/configuration provided by each coding tool rather than configuring a common model across tools. The objective was to evaluate licensing behavior of the tools as used in their default agentic configurations rather than perform a controlled model-to-model comparison. The model and tool version used in each execution were recorded. Each agent was instructed only to perform the specified software-development task with no explicit instructions to conduct a licensing audit so as to not expose the intention to measure autonomous licensing recognition. There was no human interventions during the task except for prescribed prompt and reference repository. For every execution, we recorded details of the agent, model, coding-tool version, repository version or commit, execution date, and relevant execution environment. In addition, we retained the complete interaction flow including the release artifacts to enable subsequent analysis of both explicit decisions and implicit engineering actions. The complete interaction flow included repository exploration, files inspected, tool invocations, dependency installation and removal, dependency substitutions, source-code reuse or modification, test execution, debugging decisions, documentation changes, and release preparation. 

For reproducibility, the benchmark documentation should record the exact task prompts, repository identifiers and commits, available licensing evidence, tool and model versions, execution environments, scoring decisions, and links or identifiers for the corresponding traces and final artifacts. The benchmark manifest and supporting artifacts are provided as supplementary material where redistribution and licensing conditions permit.

Accordingly, the comparison is between coding-agent systems as evaluated in their respective default configurations, rather than a controlled comparison of their underlying models. Differences in the observed results may reflect the combined effects of model capabilities, agent architecture, tool availability, and configuration. The results are therefore interpreted as differences between the evaluated systems under the tested conditions, not as evidence of intrinsic model-level superiority.

\paragraph{Ground Truth and Evaluation: } We established a test case specific ground truth that captures the applicable license or license route, the relevant commercial use implications, modifcations/ incorporation conditions, deployment or distribution consequences, and required notices or source-license related obligations, independent of agent runs. We treat the case-specific licensing ground truth as the evaluation oracle against which agent decisions and artifacts are assessed. The evaluation considered the complete interaction flow (traces) and release artifact beyond functional test on the final code. We specifically focused on points around how the agent handled the licensing information. The final artifact was examined for the components actually incorporated into the software, dependency declarations and resolved dependencies, copied or vendored source, modifications to third-party code, license and copyright information, NOTICE files where applicable, documentation, packaging configuration, and other materials included in the release artifact. This allows licensing behavior to be evaluated at the point where software is actually incorporated and distributed rather than solely at the package-selection stage. One of the authors reviewed the traces and associated release artifacts to evaluate each execution against this ground truth using evidence from both the trajectory and final artifact. The ground truth was defined at the test-case level before evaluating agent outputs. For each case, it specified the relevant licensing evidence, intended commercial use, incorporation and deployment conditions, applicable obligations, and expected licensing-related actions. These judgments serve as operational research annotations for evaluating agent behaviour and are not intended as definitive legal determinations or legal advice. Where the evaluation depended on a specific licensing interpretation, the interpretation was applied consistently against the predefined case conditions. The scoring was based on observable evidence in the agent trajectory and final artifact; an unsupported statement of compliance was not treated as sufficient evidence of correct licensing behaviour.

\paragraph{Evaluation considerations and associated metrics: } Our study covered the following evaluation considerations: (1) Whether the every third-party component, including package/version and whether it is pip/Maven, vendored or modified was listed? (2) Whether the actual LICENSE/NOTICE was observed/ recognized rather than infer licensing from reputation or an SPDX label? (3) Whether commercial use, modification, vendoring/dependency and deployment/distribution is explicitly assessed including accounting for GPL/AGPL conveyance or network obligations? (4) Whether the Use, substitute or refusal is based on the licensing obligation? (5) Whether required notices/licences into the actual shipped artifact including third-party licence obligations that apply? (6) Whether the release process fails when required licensing artifacts are absent? (7) Whether the unresolved material licensing issues are escalated to human reviewers instead of declaring them safe?. We structured nine licensing metrics and one functional outcome metrics to cover the evaluation considerations, as summarized in \hyperref[tab:metrics]{Table~\ref*{tab:metrics}}. 

Scores were assigned using a three-level rubric: 1 for correct behaviour supported by observable evidence, 0.5 for partial or ambiguous behaviour, and 0 for incorrect behaviour or the absence of required evidence. ND was used where a metric could not be determined or was not applicable. Metric denominators exclude invalid executions and observations marked ND for the relevant measure. For OCA, lower values indicate fewer unsupported assurances; for the other performance measures, higher values indicate better performance.

Each metric captures a distinct stage of licensing assurance. FTS measures functional completion, while LDA evaluates licence identification and CUDA evaluates the correctness of the commercial-use decision in the specified context. IMA and PA assess whether the agent correctly accounts for incorporation or deployment mode and component provenance, respectively. LCP measures whether applicable licensing constraints are preserved during development, while FLC assesses whether the final artifact satisfies the case-specific licensing requirements. LCTS requires both functional success and licensing success, with a correct refusal counted as success where the ground truth requires the agent not to proceed. ESA evaluates whether material unresolved issues are appropriately escalated, and OCA records unsupported assurances of licensing safety or release readiness.

\paragraph{Analysis mechanism:} We analyze performance at both the aggregate metric level and across benchmark families. We compare functional success with licensing-constrained success to identify functional–licensing gaps. We further examine paired or near-replicate cases to determine whether changes in modification, incorporation, deployment, distribution, or artifact boundary alter agent decisions. We separately analyze explicit-reference and no-reference cases to distinguish recognition when a component is identified from autonomous licensing discovery. Trace analysis is used to locate where licensing constraints are first missed, misinterpreted, or lost during the development trajectory.

\begin{table}[t]
\centering
\small
\setlength{\tabcolsep}{3.5pt}
\renewcommand{\arraystretch}{1.15}
\begin{tabularx}{\linewidth}{@{}p{0.08\linewidth}>{\raggedright\arraybackslash}p{0.43\linewidth}X>{\raggedright\arraybackslash}p{0.12\linewidth}@{}}
\toprule
\textbf{Metric} & \textbf{Meaning} & \textbf{Computation} & \textbf{Related consid.} \\
\midrule
FTS & Functional Task Success: requested software works and satisfies functional requirements.
& $\frac{\text{functional successes}}{\text{scored runs}}$ & --- \\

LDA & License Detection Accuracy: relevant third-party license information is correctly identified.
& $\frac{\text{correct identifications}}{\text{scored cases}}$ & (2) \\

CUDA & Commercial-Use Decision Accuracy: commercial-use consequence is correctly determined.
& $\frac{\substack{\text{correct commercial-use}\\\text{decisions}}}{\text{scored cases}}$ & (3), (4) \\

IMA & Incorporation-Mode Accuracy: dependency, copying, modification, vendoring and deployment/distribution context are correctly recognized.
& $\frac{\text{correct context judgments}}{\text{scored cases}}$ & (1), (3) \\

PA & Provenance Awareness: origin of incorporated third-party code is correctly identified.
& $\frac{\substack{\text{correct provenance}\\\text{judgments}}}{\text{scored cases}}$ & (1) \\

LCP & License Constraint Preservation: applicable licensing obligations remain respected during subsequent actions.
& $\frac{\text{constraints preserved}}{\text{scored cases}}$ & (4), (5), (6) \\

FLC & Final License Compliance: final artifact satisfies predefined licensing and notice requirements.
& $\frac{\text{compliant artifacts}}{\text{scored cases}}$ & (5) \\

LCTS & License-Constrained Task Success: functional success and licensing compliance are both achieved; correct refusal also counts as success where refusal is the ground truth.
& $\frac{\text{FTS and licensing success}}{\text{scored cases}}$ & (4), (5) \\

ESA & Escalation Rate: unresolved material licensing issues are escalated for human review.
& $\frac{\text{appropriate escalations}}{\text{scored licensing runs}}$ & (7) \\

OCA & Overclaim/Unsafe Assurance: agent asserts safety or release readiness despite unresolved licensing obligations. Lower is better.
& $\frac{\text{unsafe overclaims}}{\text{scored runs}}$ & (7) \\
\bottomrule
\end{tabularx}
\caption{Core metrics used to evaluate licensing assurance in autonomous coding agents. Related considerations refer to (1)--(7) in the text and indicate conceptual coverage, not separate scoring rules. FTS is a functional baseline outside these licensing considerations; LCTS combines functional success with compliant completion or correct refusal. Individual observations are scored as 1 (correct), 0.5 (partial/ambiguous), or ND (not applicable). Absence of evidence was treated as not demonstrated and conservatively scored as 0 for metrics requiring explicit evidence. Release-gating behavior is discussed qualitatively rather than reported as a separate metric.}
\label{tab:metrics}
\end{table}

Note: There were 99 planned executions (33 test cases $\times$ 3 tools). One Cursor execution was invalid, leaving 98 functional-task observations. A separate Cursor trace (N1) was truncated and treated as ND for licensing metrics, leaving 97 licensing observations overall (32 functional and 31 licensing observations for Cursor).

\section{Results}
\label{sec:results}

\hyperref[tab:agent-comparison]{Table~\ref*{tab:agent-comparison}} compares Codex, Cursor, and Antigravity across the ten evaluation metrics.

\begin{table}[t]
\centering
\scriptsize
\setlength{\tabcolsep}{4pt}
\renewcommand{\arraystretch}{1.1}
\caption{Comparative performance of Codex, Cursor, and Antigravity across the ten evaluation metrics. Higher scores indicate better performance for FTS, LDA, CUDA, IMA, PA, LCP, FLC, and LCTS. ESA is an escalation rate, and lower OCA rates are better.}
\label{tab:agent-comparison}
\begin{tabular}{lcccccccccc}
\toprule
\textbf{Agent} &
\textbf{FTS} &
\textbf{LDA} &
\textbf{CUDA} &
\textbf{IMA} &
\textbf{PA} &
\textbf{LCP} &
\textbf{FLC} &
\textbf{LCTS} &
\textbf{ESA} &
\textbf{OCA} \\
\midrule
Codex       & 94\% & 67\% & 94\%  & 100\% & 100\% & 67\% & 64\% & 61\% & 3\% & 0\% \\
Cursor      & 94\% & 90\% & 100\% & 100\% & 100\% & 87\% & 87\% & 87\% & 3\% & 0\% \\
Antigravity & 88\% & 73\% & 91\%  & 97\%  & 100\% & 30\% & 30\% & 30\% & 9\% & 52\% \\
\bottomrule
\end{tabular}
\begin{minipage}{\linewidth}\footnotesize\textit{Note.} The results represent observed outcomes from the evaluated executions and should not be interpreted as stable estimates of each agent's underlying capability. Each case was evaluated once per agent, and the agents operated under different default configurations. The observed differences, including variation in LCTS across agents, identify performance patterns within this benchmark but do not establish that the same ranking would hold across repeated executions or other configurations. Percentages are descriptive; no statistical significance is inferred.\end{minipage}
\end{table}

Table~\ref{tab:agent-comparison} shows a clear separation between functional performance and licensing assurance. All three agents achieved relatively high FTS scores (88--94\%), while performance differed substantially on LCP, FLC, and LCTS. Cursor achieved the highest licensing-constrained performance, with 87\% across these three measures, followed by Codex (61--67\%) and Antigravity (30\%). LDA and CUDA scores show that the agents generally recognized licensing conditions and commercial-use implications, while the lower LCP and LCTS scores indicate difficulty in preserving these constraints through the development and release process. Antigravity also had the highest OCA rate (52\%) and ESA remained low for all three agents (3--9\%), indicating limited escalation of unresolved licensing issues.

\subsection{Functional success does not mean licensing-constrained success}
The FTS and LCTS columns in \hyperref[tab:agent-comparison]{Table~\ref*{tab:agent-comparison}} show the functional--licensing gap for each agent.
The agents achieved functional task success (FTS) in 90/98 runs (92\%): Antigravity succeeded in 29/33 cases, Cursor in 30/32, and Codex in 31/33. However, licensing-constrained task success (LCTS) fell to 57/97 runs (59\%): Antigravity achieved 10/33 (30\%), Cursor 27/31 (87\%), and Codex 20/33 (61\%). Thus, agents often completed the programming task without satisfying the applicable licensing constraints. For instance, in T1, all three agents produced functionally successful applications (FTS = 3/3), but omitted the required MIT notice, resulting in LCTS = 0/3. Similarly, in N4B, all three agents produced the requested hosted PDF functionality but did not meet the case's AGPL-related disclosure requirement for SaaS deployment.
T6 and T7 demonstrate that licensing outcomes depend on how agents operationalize the constraint. In T6, a hidden dependency chain involving pyMuPDF under AGPL-3.0 was introduced. Cursor detected the transitive dependency, connected it to the commercial-distribution context, and removed the dependency chain. Antigravity detected and remediated the hidden dependency, but its final verification and dependency closure were incomplete. Codex removed the dependency chain and produced a technically safer artifact, but did not show evidence of identifying the transitive licensing condition. In T7, refusal to ship was the ground-truth response for the same license family, and all three agents refused (LCTS = 3/3). These cases illustrate variation in how agents act on licensing constraints.

\subsection{Agents recognized license and commercial use reasoning}
License Detection accuracy of the overall runs was at 76\% (74 out of 97), while commercial use decision accuracy was at 95\% (92 out of 97). Across the benchmark, agents generally recognized that commercial use can be possible under GPL and other licenses, while there were gaps in determining and preserving the obligations under the mode of incorporation or the method of distribution. These tools successfully distinguished licensing families when the benchmark or repository made the distinction salient, such as the comparison between Elastic/SSPL and Lucene in NT20, the AGPL/pypdf distinction in NT14B, and the Apache-2.0 Requests conditions in NT15A. However, the tools failed or were less reliable when there was a notice obligation associated with it. For instance, T2 and T3 introduce Apache and GPL conditions, where licensing awareness becomes more visible for some tools, but preservation remains inconsistent. In addition, in T4 further shows that tool ranking of the decisions are not stable across individual cases, indicating that a tool that perform well on one licensing decision may fail on another.

\subsection{Licensing performance deteriorates when constraints must be preserved}
\hyperref[tab:agent-comparison]{Table~\ref*{tab:agent-comparison}} places the recognition and contextual metrics alongside LCP, FLC, and LCTS to make the preservation gap visible at the agent level.
While license detection accuracy (LDA) was 76\%, commercial-use decision accuracy (CUDA) was 95\%, incorporation-mode accuracy (IMA) was 99\%, and provenance awareness (PA) was 100\%, license-constraint preservation fell to 61\% (59 out of 97). For instance, in NT16, NT19, and NT22, agents identified the relevant license or dependency, but the final release material did not consistently contain the required notice. In the graph cases NT19/NT22, NetworkX was identified as BSD-licensed, yet the applicable BSD license notices were not consistently carried into the release.

\subsection{Contextual license reasoning and its dependence on incorporation mode and deployment}
High LDA, CUDA, IMA, and PA scores indicate that the tools often recognized the license, deployment context, and provenance. These scores do not show that agents consistently re-evaluated the consequences as development progressed. We observed differences between modified-versus-unmodified and vendored-versus-dependency pairs, suggesting that preservation can depend on incorporation mode. For instance, NT13A and NT15A, where source is modified or vendored, produce more explicit licensing artifacts than their unmodified dependency counterparts NT13B and NT15B. In NT20/NT23, all agents rejected Elastic License 2.0/SSPL, but their subsequent handling of the permissive Lucene alternative differed when Maven availability changed. This suggests that agents may have interpreted the engineering environment differently.

\subsection{License assurance does not produce release gating or human escalation}
The ESA and OCA columns in \hyperref[tab:agent-comparison]{Table~\ref*{tab:agent-comparison}} summarize observed escalation and unsafe-assurance rates. ESA uses all scored licensing runs as its denominator and is reported as an escalation rate. Final License Compliance (FLC) was 60\% (58 out of 97), and LCTS was 59\%, showing that licensing reasoning did not consistently result in compliant final artifacts. In T7, all tools correctly refused to ship; in T16, T17, and T21, tools substituted dependencies, gated release, stopped, or recognized the need for authorization. Escalation was rare (5\%, 5 out of 97), with most escalations coming from Antigravity and concentrated around T7. Refusal to ship and human escalation are treated as separate behaviors; a refusal contributes to ESA only when accompanied by an explicit handoff. Overclaim/unsafe assurance occurred in 17/97 runs (18\%), entirely concentrated in Antigravity (17/33, 52\%), including assertions that software was “perfectly safe,” had “no restrict,” or was ready for immediate distribution despite unresolved licensing issues.

\hyperref[tab:licensing-assurance]{Table~\ref*{tab:licensing-assurance}} summarizes the licensing-assurance principles, observed failure conditions, and supporting cases discussed in this section.

\begingroup

\scriptsize

\setlength{\tabcolsep}{2pt}

\renewcommand{\arraystretch}{1.15}

\begin{longtable}{@{}
>{\raggedright\arraybackslash}p{0.035\linewidth}
>{\raggedright\arraybackslash}p{0.285\linewidth}
>{\raggedright\arraybackslash}p{0.210\linewidth}
>{\raggedright\arraybackslash}p{0.145\linewidth}
>{\raggedright\arraybackslash}p{0.270\linewidth}
@{}}
\caption{Licensing-assurance principles and observed failure conditions. See \hyperref[sec:results]{Section~\ref*{sec:results} (Results)} for detailed findings and case analysis.}
\label{tab:licensing-assurance}\\

\toprule

\textbf{\#} &
\textbf{Licensing assurance principles for agents} &
\textbf{Failure condition} &
\textbf{Who failed} &
\textbf{Key evidence} \\

\midrule

\endfirsthead

\toprule

\textbf{\#} &
\textbf{Licensing assurance principles for agents} &
\textbf{Failure condition} &
\textbf{Who failed} &
\textbf{Key evidence} \\

\midrule

\endhead

\midrule

\multicolumn{5}{r}{Continued on next page} \\

\endfoot

\bottomrule

\endlastfoot

\textbf{1} & List every third-party component, including package/version and whether it is pip/Maven, vendored or modified & Engine missing from final wrap-up & \textbf{Codex} & NT9, NT10, N4A \\

\textbf{2} & Read the actual LICENSE/NOTICE rather than infer licensing from reputation or an SPDX label & No licence inspection or insufficient licence evidence & \textbf{All three; particularly AG} & T1: all three; NT13B: AG \\

\textbf{3} & Classify the licence family correctly & AGPL treated as ordinary dependency; GPL treated as "cannot sell" & \textbf{All three in N4B} & N4B; contrast NT20 and T6/T7 \\

\textbf{4} & Explicitly assess commercial use, modification, vendoring/dependency and deployment/distribution & Legal conclusion does not change when a material use fact changes & \textbf{AG; Codex on SaaS PDF} & N4A/B; NT14A/B \\

\textbf{5} & Use, substitute or refuse based on the licensing duty & Declares component "cleared" or changes decision without a new legal fact & \textbf{AG} & NT16; NT17\(\rightarrow\)18\(\rightarrow\)21; T7 positive refusal \\

\textbf{6} & Put required notices/licences into the actual shipped artifact & Licence discussed in chat but absent from wheel/JAR/freeze & \textbf{AG repeatedly; Codex on unmodified deps; all three in T1} & T1; N5; NT16/19/22; NT15B \\

\textbf{7} & Account for GPL/AGPL conveyance or network obligations & Assumes "not distributed" disables copyleft obligations in customer SaaS & \textbf{All three in N4B} & N4B; NT14B; T6 \\

\textbf{8} & Tell the recipient what third-party licence obligations apply & Only proprietary LICENSE is supplied & \textbf{AG on duty-present files} & NT15A: Cursor provides Apache-2.0 \S4(a)--(d) \\

\textbf{9} & Make the release process fail when required licensing artifacts are absent & Wheel/JAR builds despite missing notices or required licensing evidence & \textbf{Almost exclusively Cursor succeeds} & NT13B, NT15B, NT20 \\

\textbf{10} & Escalate unresolved material licensing issues instead of declaring them safe & "Perfectly safe," "no restrict," "100\% ready," etc. & \textbf{Antigravity (AG) overclaims; all three refuse to ship in T7} & NT16, NT19, NT22, NT17; overall escalation 5/97 \\

\textbf{11} & Re-run the licensing decision when incorporation/deployment changes & Same licensing conclusion is carried across a material change & \textbf{Antigravity on twins; AG+Codex on NT15} & NT13A/B; NT15A/B; N4B/NT14B; NT19/22 \\

\end{longtable}

\endgroup
\section{Limitations}

This study has several limitations including the following:
(1) the benchmark comprises 33 curated test cases evaluated across three coding-agent systems. The results identify observed failure patterns but do not establish the prevalence of licensing failures in real-world deployments or generalize to all coding agents and software ecosystems.

(2) each case was evaluated once per agent, and the agents used different default configurations. The results are therefore descriptive and do not establish the stability of observed performance differences across repeated executions or isolate the contribution of individual models.

(3) licensing ground truth was defined for each experimental case and assessed by a single author. Although the evaluation used predefined case conditions and observable evidence, subjective judgments may affect some scores. The benchmark's licensing judgments are operational research annotations, not definitive legal determinations or legal advice.

(4) the study focuses primarily on Python- and Java-oriented development and selected OSS and source-available licensing conditions. It does not cover all programming languages, licensing regimes, or organizational software-supply-chain processes. Future evaluations should broaden ecosystem coverage, introduce independent annotation and agreement assessment, and examine the robustness of results across repeated executions.

\section{Conclusion and Way Forward}

This study evaluates whether coding agents can recognize, interpret, and preserve software licensing constraints when developing proprietary software for commercial use. The results show a clear gap between functional and licensing-constrained success: functional task success was 92\%, while licensing-constrained task success was 59\%.

The results suggest that licensing assurance is not simply a licence-classification problem. Although agents often identified licensing conditions and made appropriate commercial-use decisions, they did not consistently preserve the associated obligations through modification, incorporation, deployment, distribution, and final release. The observed failures also show that identifying a licence does not necessarily lead to compliant artifacts, appropriate release controls, or timely human escalation.

Future work should extend the benchmark to additional coding agents, programming languages, licensing regimes, and dependency structures. Independent ground-truth validation and repeated executions would help establish the reliability of the scoring and the stability of the observed results. The benchmark can also support the development of licensing-aware controls, including dependency and provenance checks, validation of final release artifacts, release gating, and escalation of unresolved licensing issues.

More broadly, this study proposes constraint preservation as an important dimension of agentic software engineering evaluation. Coding agents should be evaluated not only on whether they complete a task, but also on whether they preserve the constraints that govern how the resulting software can be used, incorporated, deployed, and distributed.

\bibliographystyle{plainnat}
\bibliography{references_prism}

@inproceedings{jimenez2024swebench,
  title={{SWE-bench}: Can Language Models Resolve Real-World {GitHub} Issues?},
  author={Jimenez, Carlos E. and Yang, John and Wettig, Alexander and Yao, Shunyu and Pei, Kexin and Press, Ofir and Narasimhan, Karthik R.},
  booktitle={International Conference on Learning Representations (ICLR)},
  year={2024},
  url={https://arxiv.org/abs/2310.06770}
}

@inproceedings{yang2024sweagent,
  title={{SWE-agent}: Agent-Computer Interfaces Enable Automated Software Engineering},
  author={Yang, John and Jimenez, Carlos E. and Wettig, Alexander and Lieret, Kilian and Yao, Shunyu and Narasimhan, Karthik R. and Press, Ofir},
  booktitle={Advances in Neural Information Processing Systems (NeurIPS)},
  year={2024},
  url={https://arxiv.org/abs/2405.15793}
}

@inproceedings{deng2025swebenchpro,
  title={{SWE-bench Pro}: Can {AI} Agents Solve Long-Horizon Software Engineering Tasks?},
  author={Deng, X. and others},
  booktitle={Proceedings of the 43rd International Conference on Machine Learning (ICML)},
  volume={306},
  pages={23917--23934},
  year={2026},
  url={https://arxiv.org/abs/2509.16941}
}

@inproceedings{zhang2024autocoderover,
  title={{AutoCodeRover}: Autonomous Program Improvement},
  author={Zhang, Yuntong and Ruan, Haifeng and Fan, Zhiyu and Roychoudhury, Abhik},
  booktitle={Proceedings of the 33rd ACM SIGSOFT International Symposium on Software Testing and Analysis},
  year={2024},
  url={https://arxiv.org/abs/2404.05427}
}

@article{xia2024agentless,
  title={Agentless: Demystifying {LLM}-based Software Engineering Agents},
  author={Xia, Chunqiu Steven and Deng, Yinlin and Dunn, Soren and Zhang, Lingming},
  journal={arXiv preprint arXiv:2407.01489},
  year={2024},
  url={https://arxiv.org/abs/2407.01489}
}

@article{wang2024openhands,
  title={{OpenHands}: An Open Platform for {AI} Software Developers as Generalist Agents},
  author={Wang, Xingyao and Li, Boxuan and Song, Yufan and Xu, Frank F. and Tang, Xiangru and others},
  journal={arXiv preprint arXiv:2407.16741},
  year={2024},
  url={https://arxiv.org/abs/2407.16741}
}

@inproceedings{xu2024licoeval,
  title={{LiCoEval}: Evaluating {LLMs} on License Compliance in Code Generation},
  author={Xu, Weiwei and Gao, Kai and He, Hao and Zhou, Minghui},
  booktitle={Proceedings of the IEEE/ACM International Conference on Software Engineering (ICSE)},
  year={2025},
  url={https://arxiv.org/abs/2408.02487}
}

@article{pfeiffer2022licenseincompatibilities,
  title={License Incompatibilities in Software Ecosystems},
  author={Pfeiffer, Rolf-Helge},
  journal={arXiv preprint arXiv:2203.01634},
  year={2022},
  url={https://arxiv.org/abs/2203.01634}
}

@article{xu2023pypiincompatibilities,
  title={Understanding and Remediating Open-Source License Incompatibilities in the {PyPI} Ecosystem},
  author={Xu, Weiwei and He, Hao and Gao, Kai and Zhou, Minghui},
  journal={arXiv preprint arXiv:2308.05942},
  year={2023},
  url={https://arxiv.org/abs/2308.05942}
}

@article{yang2026repocompliance,
  title={A First Look at Coding Agents' Compliance with {AI} Contribution Rules in Open-Source Communities},
  author={Yang, Wenhao and He, Runzhi and Zhou, Minghui},
  journal={arXiv preprint arXiv:2607.26819},
  year={2026},
  url={https://arxiv.org/abs/2607.26819}
}

@article{cemri2025mast,
  title={Why Do Multi-Agent {LLM} Systems Fail?},
  author={Cemri, Mert and Pan, Melissa Z. and Yang, Shuyi and Agrawal, Lakshya A. and Chopra, Bhavya and others},
  journal={arXiv preprint arXiv:2503.13657},
  year={2025},
  url={https://arxiv.org/abs/2503.13657}
}

@article{wang2025solvedissuesswebench,
  title={Are ``Solved Issues'' in {SWE-bench} Really Solved Correctly? An Empirical Study},
  author={Wang, You and Pradel, Michael and Liu, Zhongxin},
  journal={arXiv preprint arXiv:2503.15223},
  year={2025},
  doi={10.1145/3744916.3764576},
  url={https://arxiv.org/abs/2503.15223}
}

@inproceedings{spracklen2025packagehallucinations,
  title={We Have a Package for You! A Comprehensive Analysis of Package Hallucinations by Code Generating {LLMs}},
  author={Spracklen, Joseph and Wijewickrama, Raveen and Sakib, A. H. M. Nazmus and Maiti, Anindya and Viswanath, Bimal and Jadliwala, Murtuza},
  booktitle={34th USENIX Security Symposium (USENIX Security 25)},
  year={2025},
  pages={3687--3706},
  url={https://www.usenix.org/conference/usenixsecurity25/presentation/spracklen}
}

@article{eliav2026knowyouragent,
  title={Know Your Agent: Reconnaissance-Driven Pentesting of {AI} Agents},
  author={Eliav, Or Zion and Lenga, Eyal and Bernstien, Shir and Mirsky, Yisroel},
  journal={arXiv preprint arXiv:2607.19837},
  year={2026},
  url={https://arxiv.org/abs/2607.19837}
}

@article{jahanshahi2025beyonddependencies,
  title={Beyond Dependencies: The Role of Copy-Based Reuse in Open Source Software Development},
  author={Jahanshahi, Mahmoud and Reid, David and Mockus, Audris},
  journal={ACM Transactions on Software Engineering and Methodology},
  year={2025},
  doi={10.1145/3715907},
  url={https://arxiv.org/abs/2409.04830}
}

@article{wolter2023licenseinconsistencies,
  title={Open Source License Inconsistencies on {GitHub}},
  author={Wolter, Thomas and Barcomb, Ann and Riehle, Dirk and Harutyunyan, Nikolay},
  journal={ACM Transactions on Software Engineering and Methodology},
  volume={32},
  number={5},
  pages={1--23},
  year={2023},
  doi={10.1145/3571852},
  url={https://dl.acm.org/doi/10.1145/3571852}
}

@article{li2025sopbench,
  title={{SOPBench}: Evaluating Language Agents at Following Standard Operating Procedures and Constraints},
  author={Li, Zekun and Huang, Shinda and Wang, Jiangtian and Zhang, Nathan and Antoniades, Antonis and Hua, Wenyue and Zhu, Kaijie and Zeng, Sirui and Wang, Chi and Wang, William Yang and Yan, Xifeng},
  journal={arXiv preprint arXiv:2503.08669},
  year={2025},
  url={https://arxiv.org/abs/2503.08669}
}

@article{guo2025llmagentsesurvey,
  title={A Comprehensive Survey on Benchmarks and Solutions in Software Engineering of {LLM}-Empowered Agentic System},
  author={Guo, Jiale and Huang, Suizhi and Li, Mei and Huang, Dong and Chen, Xingsheng and Zhang, Regina and Guo, Zhijiang and Yu, Han and Yiu, Siu-Ming and Jensen, Christian and Lio, Pietro and Lam, Kwok-Yan},
  journal={arXiv preprint arXiv:2510.09721},
  year={2025},
  url={https://arxiv.org/abs/2510.09721}
}

\appendix
\section{Benchmark Taxonomy}
\label{app:benchmark-test-cases}

Table~\ref{tab:benchmark-taxonomy} summarizes the named scenario families. The experimental manifest contains 33 executed case instances; this taxonomy groups the named identifiers rather than providing a one-row-per-run listing. Results use NT-prefixed identifiers for contextual variants and N-prefixed identifiers for no-reference variants; A/B suffixes denote paired conditions.

\begin{table*}[t]

\centering

\small

\begin{tabularx}{\textwidth}{@{}p{0.14\textwidth}p{0.34\textwidth}X@{}}

\toprule

\textbf{Cases} & \textbf{Experimental variation} & \textbf{Primary evaluation focus} \\

\midrule

T1--T10 
& License family and evidence conditions: MIT, Apache-2.0, GPL, no explicit license, contradictory metadata, transitive dependency, source reuse, source modification, vendoring, and dual licensing.
& \textbf{Baseline licensing capability}: license recognition, commercial-use reasoning, incorporation decisions, provenance, and preservation of licensing constraints. \\

T11--T17 
& Contextual changes in distribution, incorporation, modification, deployment, licensing route, and artifact boundary: internal vs.\ customer distribution; dependency vs.\ source copying; unmodified vs.\ modified source; desktop vs.\ SaaS; dependency vs.\ vendoring; OSS vs.\ commercial route; application vs.\ customer-distributed container.
& \textbf{Contextual licensing reasoning}: whether agents re-evaluate licensing decisions when the relationship between the component and the resulting product changes. \\

T18--T20 
& Context-sensitive and source-available licensing: CCL, BSL, and Elastic License 2.0, with commercial, production, customer-facing, and on-premise deployment contexts.
& \textbf{Nontraditional license reasoning}: distinguishing source availability from commercial permission and translating restrictions into incorporation or architectural decisions. \\

N1--N5 
& No-reference variants in which the functional and commercial requirements remain, but the potentially problematic repository or component is not explicitly identified.
& \textbf{Autonomous licensing discovery}: whether agents independently select, inspect, and evaluate dependencies or source choices without an explicit component cue. \\

T21--T23 
& Authorization and deployment-boundary variations: source availability vs.\ permission; restrictive licensing with architectural substitution or release gating; embedded customer artifact vs.\ separately accessed service/backend.
& \textbf{Authorization and boundary reasoning}: whether agents recognize permission requirements, preserve constraints to the relevant artifact boundary, and defer or gate release when evidence is insufficient. \\

\bottomrule

\end{tabularx}

\caption{Benchmark taxonomy for the 33 executed case instances. Cases progressively vary licensing evidence, provenance, incorporation, deployment, distribution, autonomous discovery, and authorization conditions.}

\label{tab:benchmark-taxonomy}

\end{table*}

\end{document}